# Musical practice and cognitive aging: two cross-sectional studies point to phonemic fluency as a potential candidate for a use-dependent adaptation

BaptisteFauvel[1,2,3,4], Mathilde Groussard[1,2,3,4], JustineMutlu[1,2,3,4], EiderArenaza-Urquijo[1,2,3,4], FrancisEustache[1,2,3,4], BéatriceDesgranges[1,2,3,4] and HervéPlatel[1,2,3,4*]

[1]INSERM, U1077, Caen, France

[2]Université de Caen Basse-Normandie, UMR-S1077, Caen, France

[3]Ecole Pratique des Hautes Études, UMR-S1077, Caen, France

[4]CHU de Caen, U1077, Caen, France

*Correspondence and reprint requests:

Hervé Platel

Inserm - EPHE-Université de Caen/Basse-Normandie, Unité U1077

U.F.R. de Psychologie

Université de Caen/Basse-Normandie

Esplanade de la Paix

14032 Caen Cedex

France

Phone: +33 (0)2 31 56 65 91; Fax: +33 (0)2 31 56 66 93, e-mail: herve.platel@unicaen.fr

**Abstract:** Because of permanent use-dependent brain plasticity, all lifelong individuals' experiences are believed to influence the cognitive aging quality. In older individuals, both former and current musical practices have been associated with better verbal skills, visual memory, processing speed, and planning function. This work sought for an interaction between musical practice and cognitive aging by comparing musician and nonmusician individuals for two lifetimeperiods (middle and late adulthood). Long-term memory, auditory-verbal short-term memory, processing speed, nonverbal reasoning, and verbal fluencies were assessed. In Study 1, measures of processing speed and auditory-verbal short-term memoryweresignificantly better performedby musicians compared with controls, but both groups displayed the same age-related differences. For verbal fluencies, musiciansscored higherthan controls and displayeddifferent age effects.In Study2, wefoundthat *lifetime period* at training onset (childhood vs.adulthood) was associated with phonemic, but not semantic, fluencyperformances (musicians who had started to practice in adulthood did not perform better on phonemic fluency than nonmusicians). *Current frequency* of training did not account for musicians' scores on either of these two measures. These patterns of results are discussed by settingthe hypothesis of a transformative effect of musical practiceagainst a noncausal explanation.

1. Introduction

Normal brainand cognitive aging are challenging to study, as advancing age is often associated with various pathologies that compromise cognition. However, there are accumulating evidences suggesting that, even in pathology-free conditions, normal aging is associated with changes in the neural basis of cognition (Hedden and Gabrieli, 2004; Kalpouzos et al., 2009). Cognitive functions are disproportionally impacted by aging, the earliest and most concerned are processing speed, working memory, spatial ability, reasoning, and long-term memory (Salthouse, 2010; Hedden and Gabrieli, 2004). By contrast, measures of semantic knowledge and verbal abilities are relatively stable across the lifespan (Park, 2002).

Studies have reported that the brain retains its dynamic properties in aging, with structural and functional rearrangements allowing for new learning and skills improvements (Draganski and May, 2008). It has been suggested that all lifelong environmental features and demands contribute to the establishment of a cognitive reserve,thereby partly counteractingthe age-relatedcognitive decline (Foubert-Samier et al., 2010). Therefore, besides genetic dispositions, both former and current individuals' experiences mayinfluence the quality of cognitive aging (Hultsch et al., 1999; Wilson et al., 2005; Stern, 2012), raising hopes for therapeutic interventions and/or daily lifestyle recommendations (Green and Bavelier, 2008).Epidemiological works have shown that, in addition to education (Evans et al., 1993) and occupational activity (Adam et al., 2013), the degree of individuals' engagement in leisure activities covaries with cognitive functioning in old age (Wang et al., 2012). For instance, Christensen and Mackinnon (1993) found a positive statistical link between mental, social, and physical activities and the cognitive performances of a large sample of participants aged 70-89 years. Longitudinal studies have confirmed that practicing leisure activities in later life is linked to increases in cognitive reserve (Schooler and Mulatu, 2001).

Regarding musical practice,cross-sectional worksin children and young adults have reported an association with better performances across a wide range of cognitive domains.Individuals who practice music in their spare time have been found to significantly outperformmatched nonmusicianson verbal memory (Chan et al., 1998; Brandler and Rammsayer, 2003; Ho et al., 2003; Franklin et al., 2008; Jakobson et al., 2008), vocabulary (Forgeard et al., 2008), spatial ability (Brochard et al., 2004), nonverbal reasoning (Forgeard et al., 2008), short-term memory (Huntsinger and Jose, 1991; Tierney et al., 2008), and working memory (Franklin et al., 2008).

It has been hypothesized that music traininghas transformative effects on specific cognitive functions because they sharemechanisms (e.g. auditory processes involved both in music and in language; Patel and Iversen, 1997; Fauvel et al., 2013). Another claim is that music lessons, being scholar-like activities, instead of leading to specific cognitive improvements, result in a small general gain in Intelligent Quotient (IQ; Schellenberg, 2004), possiblythrough the potentiation of executive functions (Hannon and Trainor, 2007; Schellenberg and Peretz, 2008; but see Schellenberg, 2011).

However,environmental factors influencing human cognition are challenging to study as well, and previous positive results from quasi-experimental studies should be interpreted with caution regarding the direction of the causality (e.g.high-functioning people are more likely to takemusic lessons; Schellenberg, 2011). Moreover, a recent well-controlled (random assignment) longitudinal study observed that visual art or music preschool instruction did not influence children's cognitive development (Mehr et al., 2013).

In seeking for potential factors that could promote successful cognitive aging, cross-sectionalstudies have revealed that both past and recent musical training are associated with better cognitive performances in later life (Hanna-Pladdy and MacKay, 2011; Hanna-Pladdy and Gajewski, 2012). Astudy conducted in older musicians (60-83 years) with two levels of expertise (high activity (>10 years) versus low activity(<10 years) found that they scored higher than nonmusicians on the delayed recall of a geometric shape (visual long-term memory), a verbal naming task, and Parts A and B of the Trail Making Test (TMT A and B), which measures processing speed (Hanna-Pladdy and MacKay, 2011).Although the performances of the low-activity musicians were halfway between those of the controlsand high-activity musicians, suggesting a causal explanation,the differences were not significant. Moreover,those musicians who were still actively engaged in music at the time of the study did not perform better than those who had ceased.

In another study (Hanna-Pladdy and Gajewski, 2012) where general lifestyle activities were controlled for, findings were replicated for visuospatial judgment, verbal memory, verbal working memory, planning functions (nonmusicians made more rule violations in an executive function-related task), and phonemic fluency. Statistical analyses revealed that the musicians' visuospatial abilitywas predicted by recent musical engagement, whereas auditory-verbal memory seemed to be influenced by the early age of musical acquisition. The authors suggested that it could reflect differences in use-dependent adaptation periods depending on cognitive domain.

## 2. Issues

In their investigation of the association between musical practice and late-life cognition, already published workshave focused on musicians who started their training around 10 years old, whilequasi-experimentally manipulating the current engagement (active vs.inactive), and by implication, the total duration of training (more or less than 10 years; Hanna-Pladdy and

MacKay, 2011; Hanna-Pladdy and Gajewski, 2012). Therefore, the question of whether old musicianswho started practicing musicin adulthood also display better cognitive performances remains unanswered.

Moreover, any differences uncovered by comparing musicians and nonmusiciansat a single point in time are of a *quantitative* nature, and do not inform about the cognitive aging *quality* of these two populations (i.e., is musical practice associated with better performances in old age*and*weaker cognitive decline).

The first aim of the present work was to confirm that individuals practicing music since childhood display better cognitive performances thannonmusicians. Then, we wantedto explore whether these two populationsdiffer in terms of aging-related performances evolution. Finally, we looked at whether two different characteristics of practice (age at training onset and current frequency of training) account for older musicians.

Indeed, in the perspectivethat musical training influences the aging of cognitive abilities by stimulating them, we hypothesized that:

(i) Some tests' results would be higher in musically-trained participants*and*display weaker age-related differences between middle-aged and older participants compared with nonmusicians.

(ii) For cognitive functions displaying this pattern, at least one of the two characteristics of musical practice (age at onset and/or current frequency of training) would be associated with the musicians' performances.

Therefore, in Study1, we assessed several cognitive functions in middle-aged and older musicians(all having started in childhood), and compared their cognitive performances,together with the age-dependent differences, with those of middle-aged and older nonmusicians (cf. Fig. 1).

In Study 2, we wanted to explore whether two variables of practice (i.e. lifetime period at training onset and current frequency of training) could explain the oldermusicians'higher scores. We thereforeincluded a sample of older individuals who had started musiclater in life (meanage at training onset =42.7 years,*SD*=11), andmeasured their cognitive performances on tests having revealed significant interactions in Study1. In addition, we rana correlationanalysis between these test scores and all the older musicians' current frequency of training.

## 3. Study 1:Material and methods

### 3.1. Participants

This study was made according to the ethical recommendations of Helsinki agreements for human researches and cognitive investigations. Sixty-eight individuals were included.All were informed about the study's details and gave their consent for participation.

Thirty-four were amateur musicians recruited from several French conservatories or music schools (no self-educated musicians were included). They were required to have an uninterrupted time of practice of at least 4 years at the time of the study, and for a frequency of training of more than 2 hours a week during the last 6 months. To make sure that the musicians had an "active" practice, rather than a routinized one, a further inclusion requirement was that they had tohave learned at least two new music pieces in the course of the previous year. They played various musical instruments (piano, guitar, trumpet,

etc.).Overall, the mean training duration was 38.12 years ($SD$=17.7), with a mean age at training onset of 9.54 years ($SD$=4.14), and a frequency of current training of 12.6 hours per week ($SD$=10.7) (cf. Table 1). This sample was divided into a middle-aged group(19 participants, meanage =40.7 years,$SD$=9.6, min.=21, max.=55) and an oldergroup (15 participants, meanage =67.1 years,$SD$=5.2, min.=60, max.=78).

The 34 control participants were recruited from the general population. This sample was also divided into a middle-aged group (14 participants, meanage =38.86 years,$SD$=15.9, min.=22, max.=55) and an oldergroup (20 participants, meanage = 67 years,$SD$=4.4, min.=60, max.=79) (cf. Table 1).

The cut-off ages (middle-aged≤55-60≤older) were chosen because of works showing that cognitive performances fall significantly below the mean of the general population at around this age (Salthouse, 2010).

All participants were given a semi-structured questionnaire that had been specially designed for the study in order to rate demographic data (gender, age, education and occupation) (cf. Table 1). The stimulating quality of the participants' occupations was rated as follows: a value of 1 was attributed for jobless individuals; 2 for manual workers; 3 for office workers, public-sector workers, tradespeople, and storekeepers;4 for the self-employed, executive managers, and students; and 5 for physicians and company directors. The musicians were also asked to evaluate their age at training onset, and its average frequency for the last 6 months. Based on this questionnaire, the groups of participants were matched on gender, age, education, and occupational level. Moreover, middle-aged and older musician groups differed only marginally on age at training onset as well as frequency of practice (cf. Table 1).

*Table 1: Means (SDs) and inferential statistics for demographic data and characteristics of musical practice (trend/significance p values: $^\$ p\leq.1$; \*p<.05; \*\*p<.01; \*\*\*p<.001).*

| ANOVAs/Chi-squared | | | | | | |
|---|---|---|---|---|---|---|
| | Middle-aged controls (*n*=14) | Older controls (*n*=20) | Middle-aged musicians (*n*=19) | Older musicians (*n*=15) | $F/\chi^2$ | *p* |
| **Sex ratio (M/F)** | 5/9 | 7/13 | 10/9 | 8/7 | 2.15 | .54 |
| **Education (years)** | 13.6 (3.5) | 13 (2.5) | 14.6 (2.1) | 14 (2.5) | 1.3 | .28 |
| **Occupation** | 2.6 (.7) | 3.1 (1) | 3.2 (.4) | 3 (.5) | 2.2 | .1 |
| Two-sample *t*tests/Chi-square | | | | | | |
| | Middle-aged controls | Older controls | Middle-aged musicians | Older musicians | $t/\chi^2$ | *p* |
| **Sex ratio** (M/F; collapsed across age) | 12/22 | | 18/16 | | 2.15 | .14 |
| **Education** (collapsed across age) | 13.2 (2.9) | | 14.3 (2.3) | | 1.76 | .08$^\$$ |
| **Occupation** (collapsed across age) | 2.9 (.9) | | 3.1 (.4) | | 1.26 | .21 |
| **Age** (middle-aged) | 38.9 (12.9) | | 40.7 (9.7) | | -.48 | .25 |
| **Age** (older) | | 67 (4.4) | | 67.1 (5.2) | -.09 | .93 |
| **Frequency of training** | | | 15.3 (12.6) | 9.2 (6.6) | 1.71 | .1$^\$$ |
| **Age at onset of training** | | | 8.4 (3.7) | 11.2 (4.5) | -1.84 | .07$^\$$ |

### 3.2. Neuropsychological assessment

The neuropsychological assessment was made of 8 tests that are classically used in a clinical context (cf. Table 2). The entire test battery was administered in a single session which lasted about 90 minutes, and took place in a quiet room conducive to concentration. As the whole recruitment extended through 3 years and was shared between master's degree students in psychology, eight different experimenters were trained to run the tests in the most possible standardized form. In line with previous studies in the field, we focused on 6 cognitive domains(cf. Table 2).

We probed the verbal component of long-term memory using the delayed free recall of the 12 words from the Signoret BEM-144 (Signoret, 1991) (1). This test is taken from the French Batterie d'Efficience Mnésique, and requires participants to recall verbal information after a delay of approximately 7 min. Long-term memory's visual component was assessed by combining the two versions of Baddeley's Doors test (Baddeley et al., 1994), and with the delayed recall of the Rey-Osterrieth complex figure (Rey, 1959) (2). These two tests asked respectively to recognize 12 previously encoded pictures of doors, and to redraw an abstract geometrical shape copied approximately 3 min earlier. They are both standardized tests of visual memory with good face validity and available normative data. Auditory-verbal short-term memory was evaluated using the size of the forward digit span (Godefroy et al., 2008)

(3),which requires participants toimmediately recall the highest number of digits in the order they were presented. This task is also part of the Wechsler Adult IntelligenceScale-Third Edition (WAIS-III; Wechsler, 2000). Processing speed was measured with the digit-symbol coding subtest of the WAIS-III (Wechsler, 2000), and withthe number of items processed in the d2 test (Brickenkamp, 1981) (4). In these two tests, the subject is asked to associate or to discriminate visual stimuli as quickly as possible.The digit-symbol codingis part of the WAIS-III (Wechsler, 2000), and the d2 test is a valid measure of attention and processing speed based on German and American normative samples. The d2 is a time-restricted pencil-and-paper test that asks participants to cross out as many letter "d"s with two marks above or below them, in any order. The surrounding distractors are relatively similar to the target stimulus (a "p" with two marks or a "d" with one or three marks).We administered the nonverbal Raven's progressive matrices test (Wechsler, 2000)to estimateparticipants' nonverbal reasoning skills. This test is also a good proxy forSpearman's general intelligence (g) factor (5).In this test, participants are presented with an unfinished matrix of drawings, and have to choose which of the proposed answers logically completes the matrix.Finally(6),phonemic and semantic verbal fluency tasks (Cardebat et al. 1990) were administered to assess participants' verbal functions. These time-restricted tasks ask to recite as many wordsbeginning with a given letter (phonemic) or belonging to a given semantic category (semantic) as possible. They are frequently used to assess cognition after neurological damage (Henry and Phillips, 2006). Performances on these tests can be improved by applyingstrategies(clustering by- and switching between- phonemic or semantic word categories).

*Table 2: List of the tests used, thecognitive functions they assess, and the scores retained.*

| Psychometric tests | Dependent variables (outcomes) | Cognitive functions assessed |
|---|---|---|
| **Delayed recall of the Signoret BEM-144's 12 words (Signoret, 1991)** | Number of words recalled. | Verbal long-term memory, free recall. |
| **Doors test (Baddeley et al., 1994)** | Number of doors properly recognized. | Visual long-term memory, recognition. |
| **Delayed recall of Rey-Osterrieth complex figure (Rey, 1959)** | Recall fidelity (number of details, their completeness, and location). | Visual long-term memory, free recall. |
| **Forward digit span (Godefroy et al., 2008)** | Highest number of digits properly recalled in 2/3 trials. | Auditory-verbal short-term memory. |
| **Digit-symbol coding subtest of the WAIS-III (Wechsler, 2000)** | Number of digits pairedwith their proper symbol in 2 min. | Processing speed, visual scanning. |
| **d2 test (Brickenkamp, 1981)** | Sum of the number of items processed per line in 15 s. | Processing speed, visual discrimination. |
| **Semantic and phonemic fluency tasks (Cardebat et al., 1990)** | Number of words enunciated in 2 min. | Verbal functions. |
| **Raven's progressive matrices test (Wechsler, 2000)** | Number of matrices properly completed. | Nonverbal reasoning. |

### 3.3. Statistical analyses

All the statistical analyses were performed with STATISTICA software (StatSoft,(2011).STATISTICA (data analyses software), 10$^{th}$ version www.statsoft.fr).The weakest significance threshold was set at *p=.05*,uncorrected for multiple comparisons.

Two chi-square tests of independence were run to attest that the musician and nonmusician groups did not differ significantly in gender frequencies. The first compared all four groups (middle-aged controls, middle-aged musicians, older controls, and older musicians), and the second tested the musician and nonmusician groups, collapsed across age (cf. Table 1).

Two analyses of variance (ANOVAs) were conducted to attest that the four groups did not differ significantly on education and occupation level (cf. Table 1). For these variables, we also ran two *t* tests, where the musician and nonmusician groups were collapsed across age. Moreover, two-sample *t* tests were run to confirm that there was no statistical difference in age between the middle-aged musicians and nonmusicians, and the older musicians and nonmusicians. Finally, two-sample *t* tests were run on the middle-aged and older musicians' age at training onset and frequency of training (cf. Table 1).

For informative purposes, a correlation analysis was performed between all the outcome measures (cognitive scores) of all the participants (cf. Table 3).

To seek for between-group differences regarding cognitive performances, we designed two-way ANOVAs specifying each cognitive score as the dependent variable, and age (middle-aged vs.older) and group (musicians vs.nonmusicians) as independent factors. Each time, the two-way ANOVAs tested for the main effects of group and age, as well as for the Age*Group interaction effect (cf. Table 4).

*Table3: Pairwise correlations among the outcome measures (trend/significance p values: $^\$$p<.1; \*p<.05; \*\*p<.01; \*\*\*p<.001).Squares featuring tests meant to measure the same cognitive domain are shaded.*

|   | 8. Phonemic fluency | 9. Semantic fluency | 7. d2 test | 6. Digit-symbol coding | 5. Raven's matrices | 4. Forwad digit span | 3. Rey-Osterrieth complex figure | 2. Doors test |
|---|---|---|---|---|---|---|---|---|
| 1 | .43*** | .31* | .22$^\$$ | .37** | .14$^{ns}$ | .1$^{ns}$ | .18$^{ns}$ | .17$^{ns}$ |
| 2 | .1$^{ns}$ | .05$^{ns}$ | .28* | .16$^{ns}$ | .33** | .01$^{ns}$ | .38** | |
| 3 | .16$^{ns}$ | .05$^{ns}$ | .35** | .44*** | .35** | .2$^{ns}$ | | |
| 4 | .27* | .43*** | .4** | .41** | .46*** | | | |
| 5 | .19$^{ns}$ | .3* | .31* | .55*** | | | | |
| 6 | .47*** | .31* | .61*** | | | | | |
| 7 | .26* | .22$^\$$ | | | | | | |
| 9 | .48*** | | | | | | | |

1. Delayed recall of the Signoret BEM-144's 12 words; 2. Doors test; 3. Delayed recall of Rey-Osterrieth complex figure; 4. Forward digit span; 5.Raven's progressive matrices; 6.Digit-symbol coding; 7.d2 test; 8. Phonemic fluency; 9.Semantic fluency.

# 4. Study 1: Results

## 4.1. Demographic matching

The two chi-square tests of independence showed that the four groups did not differ on gender frequencies ($\chi^2$ (3, $N$=68) =2.15, $p$=.54), and nor did the musician and nonmusician groups collapsed across age, ($\chi^2$(1, $N$= 68) =2.15, $p$=.14).

No significant difference was revealed by the ANOVAs run on the mean years of education ($F$ (3, 64)=1.29, $p$=.28), and professional occupation stimulating quality ($F$ (3, 64)=2.19, $p$=.1). There was also no significant difference when comparing the musician and nonmusician groups (collapsed across age) on professional occupations ($t$ (66)=1.26, $p$=.21), but the musicians tended to be more highly educated than the nonmusicians ($t$ (66)=1.76, $p$=.08).

$T$ tests confirmed that there was no difference in age between musicians and nonmusicians, be they middle-aged ($t$ (31)=.48, $p$=.63), or older ($t$ (33)=-.09, $p$=.93).

Finally, $t$ tests comparing the middle-aged and older musicians on frequency of training ($t$ (32)=1.71, $p$=.1), and age at onset of practice ($t$(32)=-1.84, $p$=.07), revealed trends toward significance, as the middle-aged musicians tended to have started musical training earlier and to have practiced more frequently than the older musicians ($p$≤.1).

## 4.2. Between-group differences in the age effect on cognition

To explore whether musical practice interacted with the age effect, each cognitive score was entered in a two-way ANOVA, with group and age as explanatory variables (cf. Table 4).

*Table 4: Means (SDs) for neuropsychological z-scores, as well as F values resulting from the two-way ANOVAs (trend/significance p values: $^\$p<.1$; *$p<.05$; **$p<.01$; ***$p<.001$).*

|   | Middle-aged controls | Middle-aged musicians | Older controls | Older musicians | F Age | F Group | F Age*Group |
|---|---|---|---|---|---|---|---|
| 1 | -.06 (.8) | .09 (1.2) | -.14 (1) | .13 (1) | .005 | .69 | .07 |
| 2 | -.13 (1.1) | .21 (1.1) | .07 (.9) | -.24 (1) | .23 | .004 | 1.8 |
| 3 | .3 (.9) | .21 (1.1) | -.4 (.9) | .01 (1) | 3.61$^\$$ | .46 | 1.18 |
| 4 | .09 (1) | .4 (.9) | -.72 (.9) | .38 (.9) | 3.6$^\$$ | 10.43** | 3.29$^\$$ |
| 5 | .56 (.7) | .41 (.9) | -.66 (1.1) | -.16 (.6) | 17.1*** | .65 | 2.2 |
| 6 | .53 (.7) | .63 (1.2) | -.76 (.7) | -.28 (.4) | 30.3*** | 2 | .9 |
| 7 | -.23 (1) | .7 (1.12) | -.38 (.7) | -.1 (.8) | 4.1* | 6.7* | 1.92 |
| 8 | -.32 (.8) | .1 (.5) | -.5 (1.13) | .9 (.9) | 1.83 | 17.9*** | 5.2* |
| 9 | .1 (.7) | .08 (1.18) | -.5 (.9) | .48 (.8) | .18 | 4.18* | 4.67* |

1. Delayed recall of the Signoret BEM-144's 12 words; 2. Doors test; 3. Delayed recall of Rey-Osterrieth complex figure; 4. Forward digit span; 5.Raven's progressive matrices; 6.Digit-symbol coding; 7.d2 test; 8. Phonemic fluency; 9.Semantic fluency.

### 4.2.1. Verbal long-term memory

The delayed recall of the 12 words taken from the BEM-144 yielded no main effect of either group, ($F(1, 64)=.69, p=.41$), or age, ($F(1, 64)=.005, p=.94$), and no interaction effect ($F(1, 64)=.07, p=.8$).

### 4.2.2. Visual long-term memory

Performances on Baddeley's Doors test revealed no main effect ofeither group ($F(1, 64)=.004, p=.95$); or age ($F(1, 64)=.23, p=.63$); and no Group*Age interaction effect ($F(1, 64)=1.83, p=.18$). Regarding the delayed recall of the Rey-Osterrieth complex figure, results indicated no main effect of group($F(1, 64)=.46, p=.5$); a trend toward significance for the main effect of age ($F(1, 64)=3.6, p=.06$); and no interaction effect ($F(1, 64)=1.2, p=.3$).

### 4.2.3.Auditory-verbal short-term memory

The two-way ANOVA conductedon the auditory forward digit span scoresrevealed a main effect of group ($F(1, 64)=10.43, p<.01$), and trendstoward significance for the main effect of age ($F(1, 64)=3.55, p=.06$), and the interaction effect ($F(1, 64)=3.29, p=.07$).

### 4.2.4. Nonverbal reasoning

Raven's progressive matrices' scoresrevealeda significant main effect of age, $F(1, 64) = 17.1, p< .0001$, but no significant effect of group ($F(1, 64)=.65, p=.42$), or the Group*Age interaction($F(1, 64)=2.2, p=.14$).

### 4.2.5.Processing speed

Performances on the digit-symbol codingtest showed no significant group effect ($F(1, 64)=1.98, p=.16$); a significant "age" effect ($F(1, 64)=30.3; p<.001$); and no significant interaction ($F(1, 64)=.88, p=.35$).

Regarding the d2 test, analysis revealed significant effects of group, ($F(1, 64)=6.7, p<.05$), and age ($F(1, 64)=4.1, p< .05$), but no significant interaction effect ($F(1, 64)=1.93, p=.17$).

### 4.2.6. Verbal fluencies

The number of words provided during the phonemic fluency task showed a significant effect of group ($F(1, 64)=17.9$, $p<.001$), no age effect ($F(1, 64)=1.83$, $p=.18$), and a significant interaction effect ($F(1, 64)=4$, $p<.05$).

Posthoc analysis showed that the middle-aged and older controls had very similar performances ($p=.5$), whereas the older musicians performed significantly better than the middle-aged musicians ($p<.05$). It resulted in a difference in performances between the older musicians and nonmusicians ($p<.001$). The performances of the middle-aged musicians and nonmusicians were the same ($p=.18$) (cf. Fig. 2).

Regarding semantic fluency, the two-way ANOVA revealed a significant effect of group ($F(1, 64)=4.19$, $p<.05$), no effect of age ($F(1, 64)=.18$, $p=.67$), and a significant interaction ($F(1, 64)=4.67$, $p<.05$).

Posthoc analysis revealed that the older controls tended to perform worse than their middle-aged counterparts ($p=.07$), but there was no difference between the older and middle-aged musicians ($p=.22$). Again, there was a significant difference between older controls and older musicians ($p<.01$); and no difference between middle-aged controls and musicians ($p=.93$) (cf. Fig. 2).

*... Insert Figure 2 about here*

5. Study 1: Discussion

In order to explore whether regular engagement in musical activities is associated with a decrease in the age effect on some cognitive functions, we studied the cognitive performances of middle-aged and older musicians in comparison with control participants. In line with the literature, results were as follows:

5.1. No main effect of musical practice

We found no musical practice-related difference on the verbal and visual modalities of long-term memory. Hanna-Pladdy and MacKay (2011) had previously found that musical engagement in older individuals was linked to better performances in visual memory. This finding was not replicated here, despite using a virtually identical test (delayed visual figure reproduction).

We also found that musical practice was not associated with better performances on Raven's progressive matrices, suggesting that there is no link between engagement in musical activities and nonverbal reasoning skills. Correlational studies regarding musical training and nonverbal reasoning skills (assessed using virtually similar tests) have sometimes reported positive findings (e.g., Forgeard et al., 2008; Bailey and Penhune, 2012), but not always (e.g., Franklin et al., 2008; Schellenberg and Moreno, 2010). The null finding of the present study adds to the uncertainty about the association between musical training and nonverbal reasoning skills.

In the same vein, digit-symbol coding was not performed better by musician participants of our sample. Individualized music lessons (learning to play the piano) given to 16 older nonmusicians were shown to increase performances on this measure (Bugos et al., 2007). Here, we found that older musicians who had been practicing music since childhood were no better at this test.

### 5.2. An effect of musical practice that does not interact with age

For processing speed, as assessed with the d2 test, we found significantly better performances for musician individuals compared with nonmusicians, but differences between middle-aged and olderparticipants were the same for both groups. The literatureinvestigating musical practice during old age had already reporteda trendtoward significancefordifferenceon the TMTA(Hanna-Pladdy and MacKay, 2011). This finding was replicated here with a different test, and with the precision that age-related differences were the same for both musicians and nonmusicians.

Regarding the size of the digit span, we found that musicians performed significantly better than their nonmusician counterparts. This result goes againstHanna-Pladdy et al.'s two cross-sectional studies (2011, 2012),which did not report such findings even though they used the same test. In addition, we showed a trend toward significance for the interaction with age, but which was not significant. That meant, again, that the age-related differences in performances were not significantly different between the groups.

These patterns of results did not match our first hypothesis, and were likely to reflect a predisposition in musicians for learning/practicing music, rather than a stimulation of learning/practicing music. Nevertheless, it is also conceivable that features of musical practice (e.g., score reading, typing, and auditory attention) lead to an increase in neural resources for processing speed and auditory-verbal short-term memory, endowing musicians with an advantage of a quantitative nature,with no further particular change.

### 5.3. An effect of musical practice that interacts with age

For the two verbal fluency tests (phonemic and semantic), results showed that musicians produced significantly more correct words than nonmusicians. This was only partially in agreement with Hanna-Pladdy and Gajewski (2012), as they found the same effect for phonemic, but not semantic fluency. Therefore, we confirmed that older individuals who practice music score higher on a test measuring verbal skills.

In addition, significant interactions with age indicated that performances were relatively similar between middle-aged and older individuals of the control groups (with a trend toward reduced semantic fluency for older participants), but increased significantlyin musicians for phonemic fluency.

The semantic and phonemic fluency tasks were therefore the only measures that confirmed our hypothesis of different age-related changes for musicians compared with nonmusicians. However, we did not findsignificant effect of age on these measures, makingthem less interesting in the framework of cognitive reserve. Moreover, it is important to point outthat for these two tests, no difference was found between musicians and nonmusicians of the middle-age groups. This did not really suggest a transformative effect of musical practice, because in the case of an actual *stimulation*, one would have expected musicians to outperform nonmusicians from middle age onwards.

## 6. Study 2: Material and methods

### 6.1. Study 2: Participants

In this second step, we sought for a link between the characteristics of the musicians' practice and their performances on tests having revealed significant interactions in Study 1 (i.e. phonemic and semantic fluencies). To probe the age at training onset effect, we added a sample of older musicians who had started music in adulthood (called "short-term musicians"), and compared their performances with those of the age-matched participants of Study 1(the older"long-term musicians" and older controls). We also questioned the influence of a characteristic of currentpractice by running a correlation analysis between the oldermusicians' frequency of training and their performances.

Twelve older amateur musicians (mean age= 70.04 years,*SD*=7.7, min.=61, max.=83) who had started music at a mean age of 42.7 years (*SD*=11) were recruited fromFrench conservatories or music schools. As for Study 1, no self-educated musicians were included, they had learned at least 2 new pieces during the last year, and they had an uninterrupted time of practice of at least 4 years at the time of the study (meantraining duration=25.8 years,*SD*=12.38, min.=6, max.=40), with at least 4 hours' practicing a week (meanfrequency =8.08 hours,*SD*=7.24, min.=4, max.=30) (cf. Table 5).

*Table 5: Means (SDs) and inferential statistics for demographic data and practice characteristicsof short-term and long-term musicians(trend/significance p values: $^\$p<.1$; \*p<.05; \*\*p<.01; \*\*\*p<.001).*

| **ANOVAs/Chi-square** | Controls (*n*=20) | Long-term musicians (*n*=15) | Short-term musicians (*n*=12) | $F/\chi^2$ | *p* |
|---|---|---|---|---|---|
| Sex ratio (M/F) | 7/13 | 8/7 | 7/14 | 1.7 | .43 |
| Age (years) | 67 (4.4) | 67.1 (5.2) | 70 (7.7) | 1.3 | .28 |
| Education (years) | 13 (2.5) | 14 (2.5) | 14.6 (3.6) | 1.7 | .19 |
| Occupation | 3.1 (1) | 3 (.5) | 3.2 (1.1) | .4 | .7 |
| **Two-sample *t*tests** | | Long-term musicians (*n*=15) | Short-term musicians (*n*=12) | t | p |
| Training onset (age) | | 11.2 (4.5) | 42.7 (11) | -10.26 | .000\*\*\* |
| Training frequency (hours/week) | | 9.2 (6.6) | 8.1 (7.2) | .41 | .74 |

### 6.2. Study 2: neuropsychological assessment

Study 2's neuropsychological battery was exactly the same as the one used in Study1. It comprised the same tests and was administered in an identical way by the same experimenters. Zscore values were computed based on the entire Study 2's sample.

### 6.3. Study 2: statistical analyses

A Chi-Square test of independence was run on the between-group gender frequencies to attest that the male/female ratios did not differ significantly. Three ANOVAs were conducted on the meansfor age, education, and occupation, to attest that the three groups of participants were similar on these measures.

To locate the short-term musician group's performances relative to the long-term musician and the control groups, we conducted ANOVAs on the cognitive scores having revealed interaction effects inStudy 1 (i.e. the two verbal fluency tasks). Then, if the ANOVAs led to a significant effect of group, we ran posthoc analyses (Least Square DeviationLSD) for pairwise comparisons.

Finally, to see whether the musicians' frequency of training explained their number of recalled words, we conducted a nonparametric correlation analysis (Spearman's rho)between these variables for all the musicians of the Study 2's sample.

## 7. Study 2: results

### 7.1. Demographic matching

The chi-squaretest of independence showed no between-group difference for gender frequencies ($\chi^2(2, N = 47)=1.7$, $p=.43$). The one-way ANOVAs ran on age ($F(2, 44)=1.3$, $p=.28$), education ($F(2, 44)=1.71$, $p=.19$) and occupation ($F(2, 44)=.36$, $p=.7$),confirmed that thegroups did not differ on these measures (cf. Table 5).

The two-sample$t$testscomparing the practice characteristics of the two musician groups revealed significantly differentages at training onset ($t(25)=-10.26$, $p<.001$), but similar training frequencies ($t(25)=.41$, $p=.74$) (cf. Table 5).

### 7.2. Between-group differencesfor phonemic fluency

Phonemic fluency scores showed a significant main effect of group ($F(2, 44)=7.51$, $p<.05$). Posthoc analysis revealed that the long-term musicians scored significantly higher than the short-term musicians ($p<.05$),the latter being no different from control participants ($p=.46$) (cf. Fig. 3).

### 7.3. Between-group differences for semantic fluency

Semantic fluency scores revealed a significant effect of group ($F(2, 44)=6.43$, $p<.01$). Posthoc analysis indicated that there was no difference in the performances of the two groups of musicians ($p=.23$), and that the short-term musician group also performed better than control participants ($p<.05$) (cf. Fig. 3).

*... Insert Figure 3 about here*

### 7.4. Correlation analysis

Results of the correlation analysis between the musicians' semantic and phonemic fluency scores and their frequency of training revealed no significant relation (cf. Table 6).

*Table 6: Spearman's rank correlation coefficients between older musicians' frequency of training and their phonemic and semantic fluency scores.*

|  | **Phonemic fluency** | **Semantic fluency** |
|---|---|---|
| **Frequency of training (hr/week)** | .29 ($p=ns$) | .19 ($p=ns$) |

### 8. Study 2: Discussion

To seek for a link between practice characteristics and older musicians' superior performances, we conducted a second study (Study 2), in which we included a sample of different older musicians (they had begun practicing music 30 years later on average than the musicians in Study 1). Results indicated that there was no link between frequency of training and verbal fluency performances. Moreover, because the two musician groups (long-term and short-term) had identical performances on the semantic version, we concluded that the older musicians' better ability in this test was not explained by their practice, and probably constituted a confounding factor.

For phonemic fluency, results showed another pattern, indicating that musicians who had started their musical training during adulthood did not differ from controls. This could be in line with works showing that practice-related brain or cognitive differences are more pronounced when training starts at an early age (Elbert et al., 1995; Schlaug et al., 1995; Amunts et al., 1997; Steele et al., 2013). As verbal fluency assesses mainly verbal functions, our results also corroborated studies reporting better performances across a wide variety of language-related tests for children who have learned to play music (e.g. vocabulary; Forgeard et al., 2008; Moreno et al., 2009). In line with a previous investigation (Hanna-Pladdy and Gajewski, 2012), the present study showed that musical training in childhood is also associated with better literacy in old age, regardless of current musical engagement. The long-term influence of early environmental factors on brain and cognition has already been attested in animal studies (Fernández-Teruel et al., 1997). Moreover, in humans, educational attainment is known to be a major factor in determining individuals' cognitive reserve later in life (Stern, 2012), even when occupation is controlled for. In the field of music, Gooding et al. (2013) found that the amountof musical engagement from childhood until adulthood explained older individuals' verbal fluency and long-term memory performances. Therefore, it could be that the cognitive advantage we observed in the older musicianswas linkedto themusical education they had received during the sensitive period of childhood. From this perspective, music lessons would act as an environmental enrichment whichendows people who benefit it with better verbal skills in later life. Then again, given that our two studies had a cross-sectional design, our resultsmay also indicate that individualswhose verbal skills are particularly good in old age are also those who were most likelyto come to music early, and to continue their musical training lately.

## 9. General discussion

To sum up, our cognitive investigationof musiciansyielded little evidence ofreducedage-related changesowing to musicaltraining. In line with this work's hypotheses, phonemic fluency was the only cognitive variable that potentially exemplified a positive use-dependent effect inaging. Hanna-Pladdy and Gajewski (2012.) had previously reported better phonemic fluency performances inolder individuals who had practiced music in childhood, and the current status of training (active vs.inactive) was not linkedwith performances (Hanna-Pladdy and Mackay, 2011). This is in agreement with the absence of an associationwith frequency of training in the present study. In addition, we showed that the period at training onset seems to matter, as older musicians who had started musical practice in adulthood were no different from controls. This could be because there is a sensitive period for brain and cognitive plasticity (Penhune, 2011). The possibility that musicalinstruction in childhood could have a long-lasting influence on cognition in later life fits in with the definition of reserve, andthe notion that "sensitive periods are epochs in development where specific experiences have enhanced and long-lasting effects on behavior and the brain" (Knudsen, 2004).

We found that only a verbal domain-related task fulfilled our criteria to be qualified as « stimulated by musical practice in aging ». This could be viewed as anadditional argument in favor ofthe existence of a special link between musical practice and verbal skills (Chan et al., 1998; Ho et al., 2003;Forgeard et al., 2008). Explanations for this presumeduse-dependent enhancement of speech fluency include the processing of commonauditory features in music and language (Patel and Iversen, 2007), and the strengthening of the auditory-motor neural couplingthat serves both in music and in language production and perception(Bangert et al., 2006).

However, as already stated, phonemic fluency was not better performed by musicians compared with nonmusicians for the middle-aged period of life. This is not really what we would expect in the case of actual stimulation. Moreover, in this work, many factors that are known to be associated with cognitive functioning and musical engagement were not controlled for, including the early sociocultural environment (family income, parental first language, parental education, etc.; Elpus, 2013). The first step of the present study investigated older people who had an ongoing musical training since childhood untillate-life. Ecologically, such individuals are scarce, and they probably have particular personality traits that prompt them not only to play music, but also toengage in other behaviors that may help them to developa cognitive reserve (healthy diet, moderate smoking/drinking, and other physical, social, and intellectual activities). Even more important, in this study, none of the IQ scores (verbal or performance) was assessed, but they are important determinants of individuals' cognitive reserve and thelikelihoodof their taking upmusic lessons (Schellenberg, 2011). Indeed, if music lessons do lead to improvements in IQ (Schellenberg, 2004), a comprehensive work has rigorously highlighted that children with a high IQ are more inclined than children with a lower IQ to attend music lessons (Schellenberg, 2011).

In the present study, a proxy forperformance IQ (Raven's matrices test) showed no between-groups difference, but we found that the semantic fluency task was performedbetter by musicians, regardless the life period at training onset and the current frequency of training. This result could reflect a tendency of high verbal IQ people (or with a large vocabulary) to come up with- and pursue- music training later in life. However, only the musicians since childhood displayed better performances on phonemic fluency. Therefore, as in Hanna-Pladdy and Gajewski's study (2012), it could suggest that old musician individuals who started practicing music in theirchildhood display better verbal functions beyond what could be explained by their initially higher verbal IQ.

To conclude, the quality of people's cognitive reserve depends on many intricate factors. It is highly conceivable that people who keep good cognitive functioning in old age are also those who are likely to pursue or expand their daily-life activities. In turn, maintaining these activities may certainly help to *keep their cognition young*. However, inour work ($N=79/5$), which did not control for such selection bias-related confounding factors, we found that highly trained and active musicians (mean training duration = 38 years; mean frequency of training = 10.86 hours/week) did not perform any better than control participants on long-term memory (verbal and visual), nonverbal reasoning, and visual scanning (digit-symbol coding). For processing speed and auditory-verbal short-term memory, musicians scored higher but displayed no weaker age-related difference, thus ruling out a *protective effect* of practice (although both processing speed and auditory short-term memory are highly solicited in musicians). Playing music, as speaking,areauditory-motorbehaviors, and there are reports that some individuals own predisposing functional brain organization to learn faster in these abilities (Zatorre, 2013). It is therefore conceivable that some peoplekeep disposing neural substratesthat prompt them to pursue music later in life.

Only for phonemic fluencywas musical practice associated with a different age-related change in performances, depending on the period at training onset (musicians who started in adulthood performed identically to controls). Therefore, according to us, this test was the solely potential candidate for a use-dependent adaptation interacting with musicians' cognitive aging.

Acquiring an expertise through deliberate practice leads to automation and to strategic processes that allow performances to be improved in a cost-effective way (neural efficiency; Kelly and Garavan, 2005). Other styles of behavior, such as exposure to novelty, may be more efficient in increasing old individuals' cognitive reserve.

## Acknowledgements

This research was supported by a French Ministry of Research grant. We would like to thank Alice RODAIX, Isabelle RAMEAU, Elhia DEMBA, Manuella BOISSET, and Camille HOU for their help in collecting data and their investment in this project. Many thanks, too, to Mona LEBLOND for helping with the English. Finally, this paper was substantially improved during reviewing, and we would like to address our gratitude to the two anonymous referees for their crucial help.

## Conflict of interest

Authority certifies that there is no conflict of interest.

## Author and contributors

BF was actively involved in this study from design to drafting. He greatly contributed to participant recruitment and data acquisition. Finally, he conducted all the statistical analyses, produced the figures, participated in the interpretation of the results, and played a central role in revising the article in the wake of the reviewing process.

JM made a substantial contribution to this study at every stage. Her assistance was particularly invaluable during revisions, when she participated in the supplementary statistical data analyses, the interpretation of the new results, the rational organization of the ideas, and the design of the figures. Moreover, she revised the manuscript many times in order to eliminate English grammatical mistakes and improve wording.

At the outset, EA-U had provided invaluable ideas for interpreting the results. These ideas more or less vanished from the first version of the paper, but became central in the revised one. Moreover, she proved to be a huge help with the statistical data analyses. Finally, she was always generous with her advice, and shared many of her ideas during the revision process.

MG played an active part in this study from beginning to end. She helped in the design and the data acquisition coordination, and brought her theoretical knowledge to bear in the interpretation of the results. Finally, she revised the manuscript several times, allowing essential improvements to be made.

As director of the laboratory, FE ensured that we had all the material resources we needed to bring this study to a successful conclusion. His experience and knowledge in neuropsychology were essential for conceiving and designing the study, setting out the issues, and interpreting the results. Finally, he revised the manuscript, and provided relevant advice regarding the rational organization of the ideas.

BD contributed her extensive experience to this study. Her knowledge in the field of cognitive aging proved essential to the study design, the interpretation of the results, and the drafting of the manuscript. Several of her ideas lie at the very core of the final version. Even more important, she highlighted several methodological problems-and how to solve them.

HP was the person behind this whole research project, and he coordinated the teamwork from start to finish. On many occasionsduring the design of the study and the interpretation of the results, his theoretical knowledge about music cognition saved us a huge amount of time. Moreover, his knowledge regarding the state of the art helped us to exploit the data in the most relevant way.

*Figure 1: Dispersion graph showing the participants'agesin Study 1 (red square) and 2 (green square).*

*Figure 2: Phonemic and semantic fluencyz scores as a function of age for middle-aged and older musicians and controls ($^\$p<.1$; $*p<.05$; $**p<.01$; $***p<.001$).*

*Figure 3:Means and standard deviationsof phonemic and semantic fluencyz scores for older controls and older short- and long-term musicians ($*p<.05$; $**p<.01$; $***p<.001$).*